\documentclass[twocolumn,secnumarabic,showpacs,amsmath,amssymb,amsfonts,
nobibnotes, aps,
prl]{revtex4}
\usepackage{dsfont}
\usepackage{mathrsfs}
\usepackage{bbm}
\usepackage{graphicx}
\usepackage{color}

\newcommand\ONE{\mathbbm{1}}

\newcommand\rl{l}

\newcommand{\sdet}{ \mathrm{sdet} }

\begin{document}

\author{S. Gnutzmann$^1$, J.P. Keating$^2$, and F. Piotet$^2$}
\affiliation{ $^{1}$School of Mathematical Sciences,
  University of Nottingham, Nottingham NG7 2RD, UK \\
  $^{2}$Department of Mathematics, University of Bristol,
  Bristol BS8 1TW , UK}

\title{Quantum ergodicity on graphs}

\begin{abstract}
  We investigate the equidistribution of the eigenfunctions on
quantum graphs in the high-energy limit.  Our main result is
  an estimate of the deviations
  from equidistribution for large well-connected graphs.
  We use an exact field-theoretic
  expression in terms of a variant of the supersymmetric
  nonlinear $\sigma$-model. Our estimate is based on a
  saddle-point analysis of this expression and
  leads to a
  criterion for when equidistribution emerges asymptotically in the
limit of
  large graphs. Our theory predicts a rate of convergence that is a 
  significant refinement of previous estimates, long-assumed 
  to be valid for quantum chaotic systems,
  agreeing with them in some situations but not all.
  We discuss specific examples for which the theory is tested
numerically.
\end{abstract}

\maketitle 

Quantum graphs may be viewed as networks of quantum
wires or waveguides and are standard models for large molecules
and complex quantum systems. Since the seminal paper
\cite{KottosSmil} of Kottos and Smilansky, they have become a
paradigm for quantum chaos and universal interference effects
which can be observed in the statistics of their spectra and of
their eigenfunctions (see \cite{GS} for a recent review). Kottos
and Smilansky showed (numerically) that the spectral statistics
for a large class of quantum graphs follow the predictions of the
Gaussian random-matrix (RM) ensembles up to deviations which
become smaller when the size of the graph increases.  They also
derived an exact trace formula, analogous to Gutzwiller's trace
formula which is a semiclassical approximation for classically
chaotic systems. Subsequently it was found that some special
graphs, namely star graphs, exhibit intermediate spectral
statistics which do not converge to any of the RM expectations
\cite{BBK}. This discovery prompted the question as to when the
statistical properties of the spectrum are describable by means of
RM theory. This question was answered in \cite{GA}. The criterion
for RM statistics in the spectrum of quantum graphs relies on a
well-known quantum-to-classical correspondence (the corresponding
classical dynamics being a Markov process on the bonds of the
graph) and it is generically met in large well-connected graphs.
These works provide a relatively complete understanding of the
eigenvalue statistics of quantum graphs. Substantially less is
known about the eigenfunctions. In this Letter, we investigate
when the modulus of the high energy eigenfunctions is uniformly
spread (i.e.~{\it equidistributes}) over the graph -- a property
known as {\it quantum ergodicity}.

It is generally believed, and proved in many cases
\cite{Shnirelman,CdV,ZelditchZworski}, that an ergodic classical
system admits a quantum ergodic counterpart in the semiclassical
limit. Determining the rate at which the quantum ergodic limit is
approached is one of the major open problems in Quantum Chaos.
General conjectures for this rate have been put forward based on
heuristic semiclassical arguments, RM theory, the random wave
model and periodic orbit theory \cite{FP}.  Surprisingly,
establishing when quantum graphs are quantum ergodic has proved to
be a significant problem.  The issue is subtle: Neumann 
star graphs are
known not to be quantum ergodic (even though their classical
dynamics is ergodic) \cite{BKW}, but quantum ergodicity has been
proved for a family of graphs constructed from certain
one-dimensional maps \cite{BKS}. Applying the field-theoretical
approach of Gnutzmann and Altland \cite{GA} enables us to tackle
the quantum ergodicity problem for graphs, yielding a criterion
for when quantum ergodicity can be expected for  sequences of
graphs in the limit as the size tends to infinity and, most
significantly, an explicit estimate of the rate at which the limit
is approached.  Interestingly, in some situations this estimate
coincides with the expression conjectured in \cite{FP}, but in
others it differs from it in a fundamental way.

In the scattering approach a quantum graph can be defined by a
pair $(G, S)$, where $G$ is a metric graph with $B$ bonds of
finite lengths $L_b>0$ (and coordinates $0\le x_b \le L_b$), and
$S$ is a $2B\times 2B$ unitary matrix that will be discussed
below.  We assume that the graph is simple (at most one bond
between any two vertices and no loops) and connected, and that the
bond lengths are incommensurate (rationally independent) and all
in some fixed interval $0<L_{\mathrm{min}}<L_b<L_{\mathrm{max}}$.
Let us define the $2B$ double indices $\alpha=(b,d)$ for directed
bonds where $b=1,\dots,B$ refers to the bond and $d=\pm 1$ to the
two directions on each bond (the lower case greek letters $\alpha$
and $\gamma$ will be used to denote directed bonds throughout).
The matrix $S_{\alpha,\alpha'}$ contains the probability
amplitudes for a particle to be scattered from $\alpha'=(b',d')$
to $\alpha=(b,d)$.  If the end of $\alpha'$ and beginning of
$\alpha$ are different vertices then $S_{\alpha,\alpha'}=0$. We
assume that all scattering amplitudes in $S$ are
energy-independent.  A graph is time-reversal invariant if the
scattering matrix obeys $S=\left(
  \begin{smallmatrix}
    0 & \ONE_B\\
    \ONE_B &0
  \end{smallmatrix}
\right) S^T \left(
  \begin{smallmatrix}
    0 & \ONE_B\\
    \ONE_B &0
  \end{smallmatrix}\right)$ such that the scattering amplitudes
for the processes $(b,d) \to (b',d')$ and $(b', -d')\to (b,-d)$
coincide. The quantum dynamics consists in the propagation of
one-dimensional waves on the bonds of $G$, with scattering at the
vertices encoded by $S$. The complete quantum dynamics is then
encapsulated in the \textit{quantum map} defined by the unitary
matrix $U(k) = T(k) S$ where the diagonal matrix $T(k)=e^{ik L}$,
with $L=\mathrm{diag}(L_1,\dots,L_B,L_1,\dots,L_B)$, contains the
phase factors for propagation with wavenumber $k>0$ from one end
of a bond to the other. $U(k)$ describes the succession of
scattering at a vertex and propagation to the next vertex of a
wave function $\Psi(x)\equiv\{\psi_b(x_b)\}$ on the graph, where
the wave function is expressed in terms of $2B$ complex amplitudes
$a_{(b,d)}$ (which we will combine into a $2B$-dimensional vector
$\mathbf{a}$) of plane waves such that
$\psi_b(x_b)=\frac{1}{\mathcal{N}}
\left(a_{(b,+1)}e^{ik(x_b-L_b)}+a_{(b,-1)}e^{-ikx_b}\right)$
($\mathcal{N}$ is a normalization constant). The quantization
condition on a graph implies that the quantum map does not change
the wave function: $U(k) \mathbf{a} =\mathbf{a}$. Equivalently
$U(k)$ has an eigenvalue unity -- a condition that is satisfied
for a discrete set of wave numbers $k_n$ which form the
\textit{spectrum}. 
Eigenstates will be denoted by $\Psi_n$ and the corresponding
coefficients by the vector $\mathbf{a}_n$.

The classical counterpart of the quantum graph $(G,S)$ is a Markov
process on the directed bonds of $G$ which is obtained by
replacing the quantum scattering amplitudes by the classical
probabilities $M_{\alpha,\alpha'} =|U_{\alpha,\alpha'}(k)|^2 =
|S_{\alpha, \alpha'}|^2$.  The bistochastic matrix $M$ does not
depend on the wave number and propagates probability distributions
on the directed bonds \cite{KottosSmil,GS}. For a connected graph
the \textit{classical map} $M$ is generically mixing
(i.e.~$\lim_{n\rightarrow \infty}
(M^n)_{\alpha,\alpha'}=\frac{1}{2B}$) \cite{note1} and always
ergodic in the stochastic sense. The spectrum of $M$ is confined
to lie on or within the unit circle with (exactly) one eigenvalue unity which
corresponds to equidistribution on the bonds. We write the
eigenvalues of $M$ as $1-m_i$ with $m_1=0$ and $\mathrm{Re} m_i
>0$ for $i=2,\dots,2B$ -- this notation anticipates the
interpretation of the quantities $m_i$ as masses in a
field-theoretic setting: massive modes correspond to decaying
modes of the classical map while the invariant measure is
massless. The classical spectral gap
$\Delta_g=\mathrm{min}_{i=2,\dots 2B}|m_i|$ determines the slowest
decay rate in the (time averaged) classical dynamics. It will turn
out that a simple criterion for quantum ergodicity can be
expressed in terms of the spectral gap $\Delta_g$, while a more
detailed criterion uses the full set of masses $m_i$. Note, that
the topology of a graph influences the spectrum, e.g.~any graph
which can be cut into two disconnected components of similar size
by just erasing one bond has a small gap which is at most of order
$1/B$.

Our concern here is to determine how the probability density
associated with the eigenstate $\Psi_n$ is spread over $G$ as $k_n
\to \infty$. Let us introduce \textit{observables} that are
constant on each bond. These can be represented by real diagonal
$2B\times 2B$ matrices
$V=\mathrm{diag}(V_1,\dots,V_B,V_1,\dots,V_B)$. The expectation
value of the observable $V$ in an eigenstate $\Psi_n$ is given by
$$
  \langle V \rangle_n = \sum_{b} V_b \int_0^{L_b} dx_b\
  |\psi_b(x_b)|^2
  =
  \frac{\langle \mathbf{a}_n| LV |\mathbf{a}_{n}\rangle}{
    \langle \mathbf{a}_n| L |\mathbf{a}_n \rangle}+
\mathcal{O}(k^{-1})\ .
$$
With the spectral counting function $N(k)=\sum_{k_n}
\theta(k-k_n)\sim \frac{\mathrm{tr}\ L}{2\pi} k$ the spectral
average of the expectation value of the observable $V$ is given by
the metric average over the graph
\begin{equation}
  \label{LWL}
  A_V= \lim_{K\to \infty} \frac{1}{N(K)} \sum_{k_n\leq K}
  \langle V \rangle_n=\frac{\mathrm{tr}\ LV}{\mathrm{tr}\ L}\ .
\end{equation}
This statement is known as the local Weyl law and can be recovered
for any finite graphs straightforwardly, e.g.~by the
periodic-orbit approach developed in \cite{KottosSmil}. For a
uniformly distributed eigenstate the coefficients $a_\alpha$ are
all of equal modulus such that the expectation value $\langle
V\rangle_n$ coincides with the spectral mean for any observable
$V$. \textit{Quantum unique ergodicity} is the statement that such
uniform distributions are obtained in the high energy limit of any
subsequence of states $\Psi_{n_i}$ with $n_{i+1}>n_i$, i.e.~one
has $\lim_{i\rightarrow \infty} \langle V \rangle_{n_i}=A_V$. This
property turns out to be too strong for graphs and cannot be
expected to be realized because it is known that sequences of
states exist that are scarred by short periodic orbits
\cite{Scars, BKW}. The less strong property of quantum ergodicity
can be defined by the vanishing of the variance
\begin{equation}
  \label{QE}
  F_{V} =  \left(\lim_{K\to\infty} \frac{1}{N(K)} \sum_{k_{n}\leq K}
  \langle V \rangle_n^2\right)-A_V^2\ ,
\end{equation}
i.e. $F_{V} = 0$ for all observables $V$. This implies the
existence of subsequences of eigenstates of density one
for which  $\lim_{i\rightarrow \infty} \langle V \rangle_{n_i}=A_V$. 
It will be seen later, see \eqref{Meanfield},
that $F_{V}$ in general does not vanish for a finite graph. Indeed
one may expect, as for spectral statistics, that  quantum
ergodicity can only be attained in the limit $B\rightarrow
\infty$.  One therefore considers a sequence
$\lbrace(G_{\rl},S_{\rl})\rbrace$ of quantum graphs having increasing
numbers of bonds $B_{\rl+1}> B_{\rl}$. For each $\rl$, one defines
$A_{\rl,V_{\rl}}$ and $F_{\rl,V_{\rl}}$ by formulas \eqref{LWL} and \eqref{QE}
applied to $(G_{\rl},S_{\rl})$. One still has to specify the
acceptable sequences $\lbrace V_{\rl} \rbrace$ of observables. For
these it is natural to require the existence of
$\lim_{\rl\to \infty} A_{\rl,V_{\rl}}$ and a bound from above for $\vert
V_{\rl,\alpha}\vert$. Then a sequence $\lbrace(G_{\rl},S_{\rl})\rbrace$
is said to be \emph{asymptotically quantum ergodic} if $F_{\rl,V_{\rl}}\to
0$ as $\rl\to\infty$ for any acceptable sequence of observables
$\lbrace V_{\rl} \rbrace$. In order to avoid cumbersome notation, we
will separately work on a single quantum graph $(\Gamma_{\rl},
\Sigma_{\rl})$ of the sequence and drop the index $\rl$. Without loss
of generality we will also assume $A_{\rl,V_{\rl}}=0$.

Our aim is to show that a large class of such 
sequences exhibit
asymptotic quantum ergodicity and to give
explicit conditions which distinguish between ergodic and
non-ergodic sequences. In the course of our derivation we will
also obtain an estimate of $F_{V}$ for large (but finite) quantum
graphs. While previous attempts have largely relied on
periodic-orbit theory we will here employ a field-theoretic
approach. In order to implement our approach let us first
introduce
\begin{equation}
  \label{Fluct}
  \begin{split}
    \tilde{F}_V=&\lim_{K\to \infty}\frac{1}{N(K)}
    \sum_{k_n<K} \langle V \rangle_n^2
    \frac{2B \langle \mathbf{a}_n| L
|\mathbf{a}_{n}\rangle}{\mathrm{tr}\ L}
    \\
    =& \frac{4 B}{(\mathrm{tr}\ L)^2 }
    \sum_{\alpha,\alpha'=1}^{2B} V_{\alpha}L_{\alpha}
V_{\alpha'}L_{\alpha'} \xi_{\alpha \alpha'}
    \ ,
  \end{split}
\end{equation}
which vanishes asymptotically if and only if $F_V$ vanishes (note
that we assume $A_V=0$). For all practical purposes, $\tilde{F}_V$
and $F_V$ are the same for a large graph. In \eqref{Fluct} we have
implicitly defined $\xi_{\alpha \alpha'}$ which will be the
central quantity of interest. It can be expressed as
  \begin{equation} \label{xi}
  \begin{split}
    &\xi_{\alpha \alpha'} =
    \lim_{\epsilon\to 0^+ } \epsilon
    \left. \frac{d^2}{dj_+dj_-}\right|_{j_\pm=0}
    \times\\
    &\left\langle
      \frac{\mathrm{det}(\tilde{U}_\epsilon(k)+j_+
E^{\alpha'\alpha})}{
        \mathrm{det}(\tilde{U}_\epsilon(k))}
      \frac{\mathrm{det}(\tilde{U}_\epsilon(k)^\dagger+j_-
E^{\alpha\alpha'})}{
        \mathrm{det}(\tilde{U}_\epsilon(k)^\dagger)}
    \right\rangle_{k} ,
    \end{split}
  \end{equation}
where $ \langle \ldots \rangle_{k} = \lim_{K\to\infty} \frac{1}{K}
\int_{0}^K \ldots dk$,
$\tilde{U}_\epsilon(k)=\ONE-e^{-\epsilon}U(k)$ and $E^{\alpha'
\alpha}_{\gamma' \gamma}=\delta_{\gamma' \alpha'} \delta_{\alpha
\gamma}$.

Using a standard procedure, $\xi_{\alpha \alpha'}$ can be
expressed exactly in terms of a variant of the supersymmetric
nonlinear $\sigma$-model \cite{GA}. The procedure has two main
ingredients: (\textit{i}) 
for incommensurate bond 
lengths 
the
integral over $k$ in \eqref{xi} can be replaced exactly by an
integral over $B$ independent phases \cite{BarraGaspard},
(\textit{ii}) Zirnbauer's color-flavor transformation
\cite{Zirnbauer}. For broken time-reversal symmetry the
$\sigma$-model for the graph leads to the exact expression
\begin{equation}
  \label{sigma_ex}
  \begin{split}
    &\xi_{\alpha \alpha'} =
    \lim_{\epsilon\to 0^+ } \epsilon
    \left. \frac{d^2}{dj_+dj_-}\right|_{j_\pm=0}
    \times\\
    &\int d(Z,\tilde{Z})\,
    \sdet(\ONE- Z \tilde{Z})\,
    \sdet^{-1}(\ONE- Z \mathcal{S}_+ \tilde{Z} \mathcal{S}_-).
  \end{split}
\end{equation}
Here $Z=\mathrm{diag}(Z_1,\dots,Z_B)$ and
$\tilde{Z}=\mathrm{diag}( \tilde{Z}_1,\dots, \tilde{Z}_B)$ are
$4B\times 4B$ block-diagonal supermatrices where each block is a
$4\times 4$ matrix $Z_{b, dd', ss'}$ ($d=\pm1$ is the direction
index and $s=\mathbf{B},\mathbf{F}$ refers to bosonic and
fermionic sectors of the supermatrix, see \cite{GA} for further
details). The matrices $\mathcal{S}_+=(\ONE-j_+ E^{\alpha' \alpha}
P_{\mathbf{FF}}) S$ and $\mathcal{S}_-^\dagger=S^\dagger(\ONE-j_-
E^{\alpha \alpha'} P_{\mathbf{FF}})$ (where $P_\mathbf{FF}$ is the
projector onto the fermionic sector) contain the scattering
amplitudes and source terms $\propto j_\pm$. For time-reversal
symmetric graphs an exact $\sigma$-model can be derived with
slightly more technical effort. The result looks formally similar
to \eqref{sigma_ex} with the superdeterminants replaced by their
square roots and with the dimension of the matrices doubled (see
\cite{GA}). Here we will give explicit formulae only for broken
time-reversal symmetry but will state the final results also for
the time-reversal invariant case. More details will be given
elsewhere \cite{longversion}.

We calculate the integral over $Z$ and $\tilde{Z}$ in a
saddle-point approach (writing $\sdet^{\pm
1}(\ONE-ZA\tilde{Z}B)=\exp(\pm\mathrm{str}\,
\log(\ONE-ZA\tilde{Z}B))$) where the saddle-point analysis is
performed at $j_\pm=0$. The saddle-point equations lead to a
mean-field solution $Z_{b, dd', ss'}= Y_{ss'} \delta_{dd'}$,
$\tilde{Z}_{b, dd', ss'}= \tilde{Y}_{ss'}\delta_{dd'}$ where the
$2\times 2$ matrices $Y$ and $\tilde{Y}$ span the saddle-point
manifold. Note that the dependence of the integrand in
\eqref{sigma_ex} on the scattering matrix $S$ of the graph drops
out for mean-field configurations so that these give a universal
contribution to $\xi_{\alpha \alpha'}$. Configurations orthogonal
to the saddle-point manifold are taken into account in a Gaussian
approximation -- here, crucially, the system dependence does not
drop out. There is a direct correspondence between the modes of
the classical map $M$ and the field configurations.
Equidistribution on the graph corresponds to the unique eigenvalue
unity of $M$ on one side and to the mean-field configuration on
the other side. This is the massless mode which is calculated
exactly in our approach. The decaying modes of the classical map
correspond to the (massive) configurations of the supermatrices
$Z$ and $\tilde{Z}$ which are orthogonal to the saddle-point
manifold. The contribution of the massive modes (and their
coupling to the mean-field mode) are only taken into account
approximately. This goes beyond the diagonal approximation in a
periodic-orbit approach to $\xi_{\alpha \alpha'}$, which is
equivalent to a Gaussian approximation of \textit{all} modes in
the field-theoretic approach and which actually diverges like
$1/\epsilon$ for the quantity we are interested in (further details
will be published
elsewhere \cite{longversion}).

The contributions from the
mean-field configurations and the Gaussian fluctuations 
turn out to be
additive \cite{note2} 
and thus we may write the fluctuations as
$\tilde{F}_V=\tilde{F}_V^{\mathrm{meanfield}}+
\tilde{F}_V^{\mathrm{Gauss}}$. An explicit calculation yields
\begin{equation}\label{Meanfield}
  \tilde{F}_V^{\mathrm{meanfield}}=\beta
  \frac{\mathrm{tr} L^2V^2}{(\mathrm{tr} L)^2}
\end{equation}
for the universal contribution. Here $\beta=1$ for broken
time-reversal invariance and $\beta=2$ for time-reversal invariant
graphs.
The integration over the Gaussian fluctuation yields
\begin{equation} \label{2ndorder}
    \tilde{F}_{V}^{\mathrm{Gauss}} =
    \beta
    \frac{2}{(\mathrm{tr} L)^2}\mathrm{tr}' \frac{M}{\ONE-M}L^2V^2
\end{equation}
where the trace is over all massive modes such that
the unit eigenvalue of $M$ is excluded.

\begin{figure}
\includegraphics[width=0.42\textwidth]{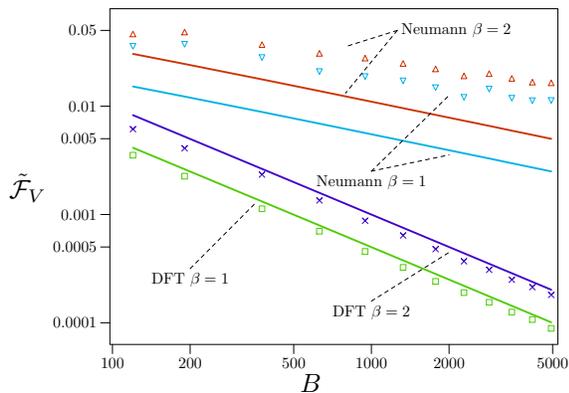}
\caption{ (color online) Deviations $\tilde{F}_V$
from quantum ergodicity for sequences of
complete graphs.
Data points refer to numerical calculations and lines refer to the
saddle-point approximation. From top to bottom: complete Neumann
graph ($\beta=2$, red), complete Neumann graph with magnetic field
($\beta=1$, light blue),
 complete DFT graph ($\beta=2$, dark blue),
 complete DFT graph with magnetic field
 ($\beta=1$, green).
}
\end{figure}

Equations \eqref{Meanfield} and \eqref{2ndorder} are our main
results. Crucially, \eqref{2ndorder} shows that the behavior of
$\tilde{F}_V$ for large graphs depends on the spectrum of the
classical map $M$. Slowly decaying classical modes with
eigenvalues near $1$ give rise to large deviations from
universality in the rate of quantum ergodicity.  This does not
conform to expectations based on the conjectures in \cite{FP},
which correspond to ignoring the Gaussian fluctuations.

In the remainder we discuss the different behaviors that sequences
of quantum graph may exhibit depending on the spectral properties
of $M$ and on the range of validity of equations
\eqref{Meanfield} and \eqref{2ndorder}. The dependence of
$\tilde{F}_V$ on the observable $V$ is of minor importance here
and we restrict ourselves to the most simple generic choice, $V_b
L_b= \pm \textrm{const}= \pm \frac{\mathrm{tr}\ L}{2B}$ such that
\begin{equation}\label{genV}
  \tilde{F}_V^{\mathrm{meanfield}}=
  \frac{\beta}{2B}
  \qquad
  \tilde{F}_{V}^{\mathrm{Gauss}} =
  \frac{\beta}{2B^2}\sum_{i=2}^{2B}
  \frac{1-m_i}{m_i}.
\end{equation}
Obviously the universal part vanishes in the limit $B\rightarrow
\infty$ without any conditions on the graph. 
We 
conjecture that the sequence of graphs is asymptotically quantum 
ergodic if and only if $\tilde{F}_{V}^{\mathrm{Gauss}} $ also
vanishes in this limit.
If the spectral gap
remains finite as $B\rightarrow \infty$ the contribution
from the massive modes remains at most of same order as the
mean-field contribution. If the spectral gap decreases as
$B\rightarrow \infty$ we may assume $\Delta_g \sim B^{-\alpha}$
with $\alpha\ge 0$. The contribution from the Gaussian
fluctuations can 
then be bounded by $\left| 
\tilde{F}_{V}^{\mathrm{Gauss}} 
\right| \leq \frac{ \beta}{B\Delta_g}\sim B^{\alpha-1}$. Based on
the saddle-point approximation we can thus give the following
criteria: (\textit{i}) a sequence of quantum graphs is
asymptotically quantum ergodic if $B \Delta_g \rightarrow \infty$
i.e. $\alpha<1$, (\textit{ii}) a sequence of quantum graphs is
\textit{not} asymptotically quantum ergodic if $\alpha\ge 2$,
(\textit{iii}) if $1\le \alpha <2$ asymptotic quantum ergodicity
depends on all masses (i.e.~on the complete spectrum of the
classical map).

Finally, let us comment on the validity of the saddle-point
approximation in this context. The masses $m_i$ determine the
stability of the saddle-point manifold -- for a single mode the
Gaussian approximation can be expected to capture the dominant
contribution if $m_i \gg 1/B$. When this condition is not
satisfied the saddle-point approximation is not valid. However,
even if $m_i\gg1/B$ the number of modes still increases
proportionally to $B$, which makes it hard to estimate the error.
Especially if there are many small masses one may expect
additional contributions to $\tilde{F}_V$ which may be calculated
by going beyond the Gaussian approximation. Thus while we believe
that our criterion for quantum ergodicity $\Delta_g B \rightarrow
\infty$ is correct we do not expect the Gaussian approximation to
give the exact asymptotic result if $\tilde{F}_V$ is dominated by
the massive contributions. We have checked numerically the
validity of the saddle-point approximation for two sequences of
complete graphs with different scattering matrices; see Fig. 1. In
the first sequence we have used discrete Fourier transform (DFT)
matrices at each vertex. In order to break time-reversal
invariance we also added magnetic fields. In this case
$\tilde{F}_V^{\mathrm{meanfield}}\gg
\tilde{F}_V^{\mathrm{Gauss}}\approx0$ and the numerics show
perfect agreement. In the second sequence we have used a
scattering matrix which is equivalent to the so-called Neumann (or
Kirchhoff) boundary conditions at each vertex. These strongly
enhance backscattering which leads to a large number of small
masses. As one may expect, the saddle-point approximation is far
from perfect in this case but still gives a reasonable estimate of
the rate.

We would like to thank Uzy Smilansky for invaluable discussions.


\begin{thebibliography}{10}
\bibitem{note1} There are only a few exceptions: e.g. the classical
map
  for a bipartite graph is not mixing.
  However, one can define a reduced quantum map and a corresponding reduced
  classical map which is mixing.
\bibitem{note2}
Note that for $j_\pm \neq 0$ the saddle-point manifold
becomes coupled to the Gaussian fluctuations. This coupling turns
out to be essential to obtain a contribution to $\xi_{\alpha
\alpha'}$ in the limit $\epsilon \rightarrow 0$. The details 
will be discussed in \cite{longversion}.
\end{thebibliography}
\end{document}